\documentclass[%
    twocolumn,
    superscriptaddress,
    bitnotes,
    amsmath, amssymb, 
    aps, 
    prl,
]{revtex4-2}

\usepackage{ulem}
\usepackage{graphicx}       % Include figure files
\usepackage{dcolumn}        % Align table table columns on decimal point
\usepackage{bm}             % Bold math
\usepackage{color}
\usepackage{subfigure}
\usepackage{appendix}
\usepackage{ulem}
\usepackage[utf8]{inputenc}
\usepackage{comment}
\usepackage{hyperref}
\hypersetup{
    colorlinks=true,
    linkcolor=blue,
    filecolor=magenta,      
    urlcolor=cyan,
    citecolor=ROYALPURPLE,
    pdfpagemode=FullScreen,
    }

\newcommand{\tcb}{\textcolor{blue}}

\definecolor{ROYALBLUE}{rgb}{0.25, 0.41, 0.88}
\definecolor{ROYALPURPLE}{rgb}{0.8, 0.3, 0.66}

\begin{document}

%\title{Unsupervised detection of continuous quantum phase transition from single basis measurements}
%\title{Unsupervised detection of off-diagonal order from diagonal basis measurements}
\title{Extracting Off-Diagonal Order from Diagonal Basis Measurements}

\author{Bo Xiao}
\email{bxiao@flatironinstitute.org}
\affiliation{Center for Computational Quantum Physics, Flatiron Institute, 162 Fifth Avenue, New York, New York 10010 USA}

\author{Javier Robledo Moreno}
\email{jrm874@nyu.edu}
\affiliation{Center for Computational Quantum Physics, Flatiron Institute, 162 Fifth Avenue, New York, New York 10010 USA}
\affiliation{Department of Physics, New York University, New York, New York 10003, USA}

\author{Matthew Fishman}
\affiliation{Center for Computational Quantum Physics, Flatiron Institute, 162 Fifth Avenue, New York, New York 10010 USA}

\author{Dries Sels}
\affiliation{Center for Computational Quantum Physics, Flatiron Institute, 162 Fifth Avenue, New York, New York 10010 USA}
\affiliation{Department of Physics, New York University, New York, New York 10003, USA}

\author{Ehsan Khatami} 
\affiliation{Department of Physics and Astronomy, San Jos\'{e} State University, San Jos\'{e}, CA 95192 USA}

\author{Richard Scalettar}
\affiliation{Department of Physics and Astronomy, University of California, Davis, CA 95616 USA}

\date{\today}

\begin{abstract}
Quantum gas microscopy has developed into a powerful tool to explore strongly correlated quantum systems.  However, discerning phases with topological or off-diagonal long range order requires the ability to extract these correlations from site-resolved measurements.  Here, we show that a multi-scale complexity measure can pinpoint the transition to and from the bond ordered wave phase of the one-dimensional extended Hubbard model with an off-diagonal order parameter, sandwiched between diagonal charge and spin density wave phases, using only diagonal descriptors. We study the model directly in the thermodynamic limit using the recently developed variational uniform matrix product states algorithm, and draw our samples from degenerate ground states related by global spin rotations, emulating the projective measurements that are accessible in experiments. Our results will have important implications for the study of exotic phases using optical lattice experiments.
\end{abstract}

\maketitle
%\section{\label{sec:intro} INTRODUCTION}
\paragraph{Introduction.--}
% Machine learning (ML) is %%largely perceived as 
% one of the main disruptive technologies in the past decade.  ML has been widely applied across many domains in physics, including statistical physics, particle physics, cosmology, quantum many-body physics and quantum computing to gain new insights and solve forefront problems 
% %% in recent years 
% ~\cite{Metha2019, Carleo2019}.  Physics, in turn, has played a key role in the development
% of ML technology\cite{karniadakis2021physics}.  In the realm of quantum many-body physics, ML approaches are employed to construct novel representation of quantum states \cite{Carleo2017, Choo2019J1J2, Hibat2020Recurrent, Nomura2017slaterRBM, Luo2019backflow, Pfau2020Ferminet, Hermann2020Paulinet, Choo2020chemistry, Robledo2021Hidden},
% %% .  ML is increasingly being used as 
% enabling it as a tool to understand the appearance of complex patterns in
% %% a method to analyze quantum states of matter, especially in systems like 
% topological materials, exotic superconductors, and spin liquids, where there are many competing types of order
% %% , and hence the appearance of complex patterns~
% \cite{schleder2019dft,stanev2021artificial}.  

Quantum gas microscopy for ultracold atoms in optical lattices, in which high-resolution real-space snapshots of the many-body system are accessible, is a prominent tool for studying strongly-correlated systems ~\cite{Sherson2010, Bakr2010, Gross2021}. 
% Experimental data is sometimes more constrained in which descriptors are available.
% In such situations, a crucial question is the extent to which 
% cleverly chosen artificial intelligence (AI) algorithms can be developed to reveal order
% from input data which are not the most natural representation of the underlying phases.
% %% Our work has an important relevance to a range of experimental
% %% methodologies and their potential use in extracting the physics of model
% %% Hamiltonians.  
% For example, quantum gas microscopy commonly provides
% real space images of optical lattices emulators of fundamental condensed
% matter model systems~\cite{Bakr2009}.  
These projective measurements can be analyzed `by hand' with
traditional counting to compute observables, both local or extended spin and charge
correlations~\cite{m_parsons_16,l_cheuk_16,p_brown_17,a_mazurenko_17,t_hartke_22}.  The snapshots are often termed `diagonal' 
since they comprise measurements of density observables $n_{i\sigma} = 
\langle c_{i\sigma}^\dagger c_{i\sigma}^{\phantom{\dagger}}\rangle$, where $c_{i\sigma}^\dagger$ ($c_{i\sigma}^{\phantom{\dagger}}$) is the spin-$\sigma$ fermion creation (destruction) operator at site $i$, 
which have matching row and column indices of the Greens function $G_{ii}^{\sigma}$.

The same is true for the outcome of large-scale programmable quantum simulators based on Rydberg atoms, which allow arranging a large number of qubits in arbitrary lattice geometries and controlling the Hamiltonian evolution of the system~\cite{Schauss2012, Ebadi2021, Scholl2021, Samajdar2020, Browaeys2020, Labuhn2016, Bernien2017}.
%However, as ML is increasingly employed to characterize more
%subtle aspects of the physics of ultracold atomic gases~\cite{Torlai2018,Khatami2020}, 
A crucial open question is whether the fact that these experiments do not at present capture `off-diagonal' information encoded in the
full $G_{ij}^\sigma$ will limit the insight they can yield.

Recent advances in machine learning methods %and their broad applications across many domains in physics, including statistical physics, particle physics, cosmology, quantum many-body physics, and quantum computing
\cite{Metha2019, Carleo2019,johnston2022perspective} hold promise for answering this question. Convolutional neural networks and hybrid supervised-unsupervised approaches have been used to classify quantum gas microscopy data in emulations of the two-dimensional Fermi-Hubbard model~\cite{Bohrdt2019}, to visualize and identify multi-particle diagonal correlations~\cite{Khatami2020, Miles2021Correlator}, and to detect new diagonal ordered phases in Rydberg atom quantum simulators~\cite{Miles2021Rydberg}. Momentum-space images of cold atoms have also been analyzed to identify quantum phase transitions~\cite{Rem2019, Kaming2021}.  

Machine learning methods are able to capture order parameters or relevant thermodynamic quantities in classical as well as quantum systems, and therefore, detect symmetry-breaking phases~\cite{Wang2016,Carrasquilla2017,Broecker2017,Chng2017,Zhang2017,Hu2017,Van2017,Wetzel2017}.  In contrast, it is much harder to identify topological phase transitions involving off-diagonal long range order.  In the realm of classical statistical physics, the two-dimensional XY and q-state clock models have been investigated to identify Berenzinskii-Kosterlitz-Thouless (BKT) transitions~\cite{Beach2018, Rodriguez2019, Miyajima2021}.  However, much less is known for BKT-type {\it quantum} phase transitions.

\begin{figure*}
    \centering 
    \includegraphics[width = \textwidth]{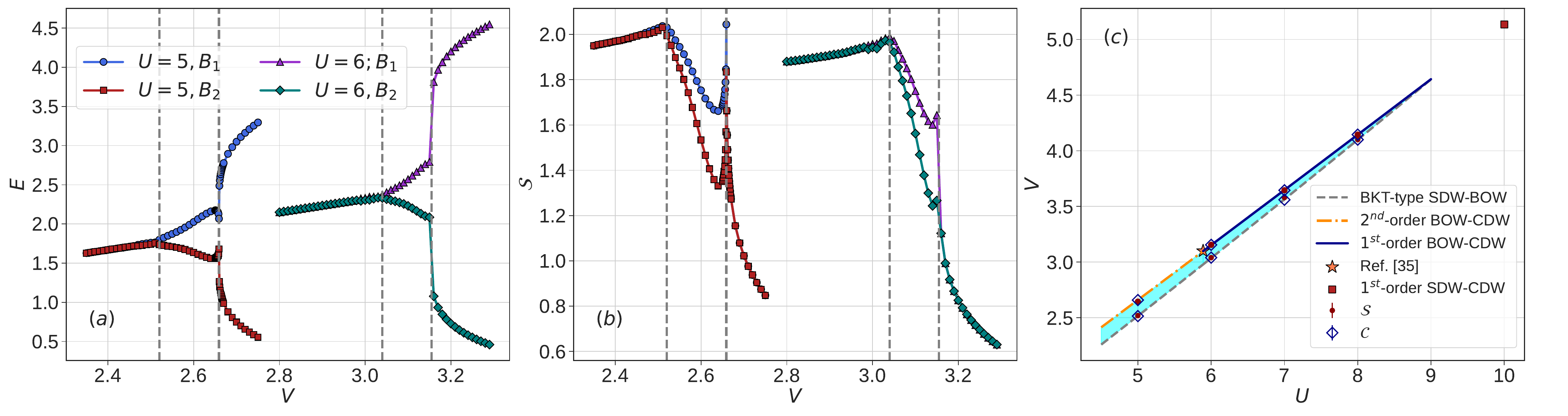}
    \caption{
    % (a) Energy $E$ and (b) von Neumann entanglement entropy $\mathcal{S}$ obtained via partitioning along two bonds $B_1$ and $B_2$ associated with one two-site unit cell vs.~nearest-neighbor interaction $V$ at fixed $U = 5$ and $U = 6$.
    (a) Energy $E_{i}$ on two bonds $B_{i} (i = 1, 2)$ associated with one two-site unit cell and (b) von Neumann entanglement entropy $\mathcal{S}_{i}$ computed via partitioning along these two bonds vs.~nearest-neighbor interaction $V$ at fixed $U = 5$ and $U = 6$.  (c) Phase diagram of the one-dimensional extended Hubbard model at half filling in the $U-V$ plane.  Phase boundaries are determined using the structural complexity $\mathcal{C}$ (blue diamond) and corroborated by the entanglement entropy $\mathcal{S}$ (red circle) at select $U$'s.  The blue shaded region denotes the BOW phase which is characterized by off-diagonal long-range order.  Orange star indicates a tri-critical point where the nature of BOW-CDW transition switches from second-order to first-order, based on Ref~\cite{Ejima2007}.  The red square indicates a direct first-order transition from SDW to CDW at $U=10$, determined using the structural complexity $\mathcal{C}$.  Bond dimension $\mathcal{D} = 2000$ is employed.}
    \label{fig:fig1}
\end{figure*}

The simplest context in which this issue can be explored is that of quasi-one-dimensional materials, e.g.~organic conductors, carbon nanotubes~\cite{Ishiguro1990, dagotto1996surprises, hu1999chemistry, Ishii2003, Baier2016}, for which the one-dimensional extended Hubbard model is a minimal description~\cite{shiba1972magnetic, imada1989numerical, schulz1990correlation, Sengupta2002, clay2003pattern, essler2005one}:
\begin{eqnarray}
    \begin{aligned}
        H = &-t \sum_{i, \sigma} \left(c_{i, \sigma}^{\dag}c_{i+1, \sigma}^{\phantom{\dagger}} 
        + c_{i + 1, \sigma}^{\dag}c_{i, \sigma}^{\phantom{\dagger}} \right)  \\
        &+ U \sum_{i} n_{i, \uparrow} n_{i, \downarrow} 
        + V \sum_{i, \sigma \sigma^{\prime}} n_{i, \sigma} n_{i+1, \sigma^{\prime}}
        \hskip0.07in , 
    \end{aligned}
\end{eqnarray}
where $U$ and $V$ are on-site and nearest-neighbor Coulomb interactions and $t=1$ sets the unit of energy.  An infinitesimally small $U$ drives a transition to a 
%\sout{spin density wave (SDW) phase in which up and down spin fermions alternate on neighboring sites}
regime with quasi-long range spin order ~\cite{essler2005one}. We will refer to this as a `SDW', but emphasize that the ground state spin correlations decay as a power law.
Similarly, an infinitesimally small $V$ induces charge density wave (CDW) order with staggered empty and doubly occupied sites~\cite{essler2005one}.  However, much less obvious is the existence, between these two phases, of a narrow bond ordered wave (BOW) region with alternating large
and small kinetic energy on adjacent sites, and a BKT-type transition separating it from the SDW phase~\cite{Sandvik1999, Nakamura2000, Tsuchiizu2002, Sengupta2002, Jeckelmann2002, Sandvik2004, Zhang2004, Ejima2007} (see Fig.~\ref{fig:fig1} (c)).  Converged results on the exact location of this BKT-type transition have not been obtained~\cite{Ejima2016, Spalding2019, Dalmonte2015, Julia2022}.  The model thus offers a unique opportunity to test machine learning tools for examining subtle quantum phase transitions characterized by non-diagonal order.

In this paper, we use the state-of-the-art variational uniform matrix product states (VUMPS) algorithm~\cite{Zauner-Stauber2018, Zauner-Stauber2018topological, Vanderstraeten2019} to obtain the ground state of the model at half filling, 
% \tcg{[MF: Weren't there iDMRG studies?]}, 
directly in the thermodynamic limit (TDL). We then emulate projective and diagonal measurements on optical lattice experiments by sampling spin-resolved occupancy snapshots from the VUMPS wavefunction. These snapshots are first analysed using principal component analysis (PCA), and then using a recently proposed {\it structural complexity} measure~\cite{Bagrov20, Sotnikov22}. We find that while PCA accurately captures the first and second order transitions between the BOW and CDW phases and the associated CDW order parameter, it fails to identify differences between the SDW and BOW samples. The structural complexity, on the other hand, starts off with a long {\it bitstring} consisting of concatenated samples and through a series of coarse-graining steps, is able to deduce the location of both transitions.

\paragraph{Variational uniform matrix product states.--}

Inspired by tangent space ideas~\cite{Haegeman2011, Haegeman2016, Zauner-Stauber2018}, VUMPS optimizes a translational invariant matrix product state (MPS) directly in the TDL, in contrast to the more traditional infinite size density matrix renormalization group (iDMRG)~\cite{McCulloch2008, White1992, Schollwock2005}
% [\tcg{MF: Cite https://arxiv.org/abs/0804.2509 and original DMRG papers by Steve White}] 
algorithm which starts from a small system and grows the state one site at a time.  

\begin{figure*}
    \hspace{-.2in}
    \centering
    \includegraphics[width = 1.\textwidth]{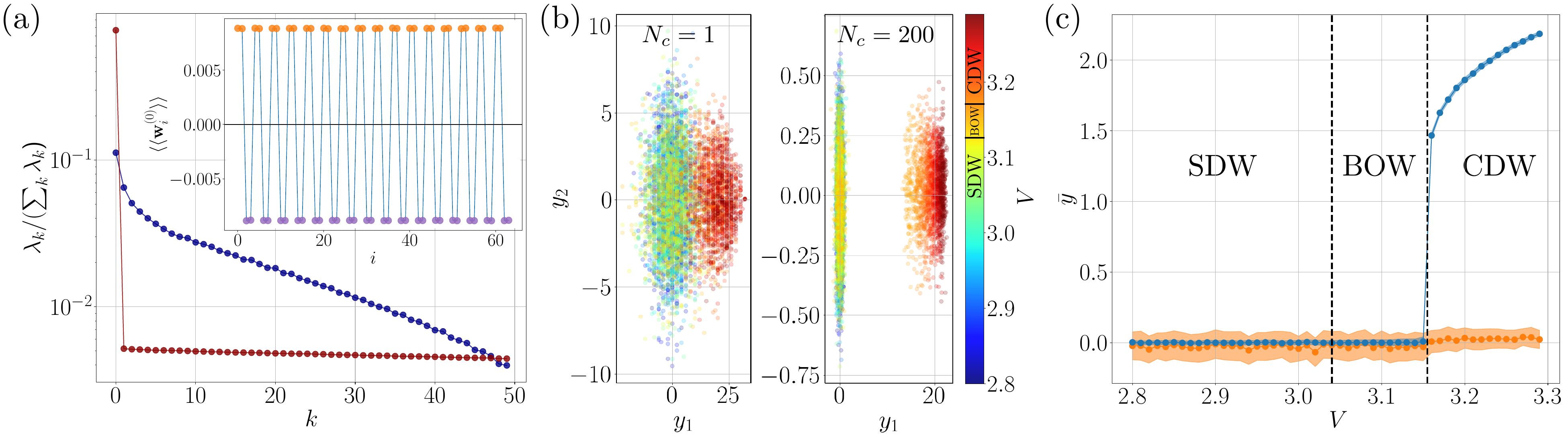}
    \caption{Principal component analysis of the samples generated at $U = 6$ and different values of $V \in [2.8, 3.3]$ in a uniform grid with separation $\Delta V = 0.1$. (a) 
    % \sout{Explained variance ratio measured by } 
    The relative weight of the eigenvalues of the covariance matrix for two values of concatenated samples $N_\textrm{c}$ \tcb{(see~\cite{SM})}. The inset shows the amplitude of the first principal component for $N_{\textrm{c} }= 200$, averaged over the concatenated samples. (b) Projection of the input features to the first two principal components both when $N_{\textrm{c} }= 1$  and $N_{\textrm{c} }= 200$. The color indicates the value of $V$ to which each projected feature belongs. (c) Average projection of the input features corresponding to a fixed $V$ to the first (blue) and second (orange) principal components for $N_{\textrm{c} }= 200$. The shaded region shows the variance of the average, measuring the spread of the projected input features at a fixed $V$. Vertical dashed lines indicate the phase transition points obtained form the entanglement entropy.}
    \label{fig:fig2}
\end{figure*}

Similar to DMRG, the energy minimization problem is reformulated as a series of local eigenvalue problems of effective Hamiltonians projected into the MPS basis. In practice a linear solver is used to perform the sum of the formally infinite number of Hamiltonian terms to obtain the effective Hamiltonians.
By working directly with a translational invariant ansatz in VUMPS, we can remove the solitonic excitation induced by the use of open boundary conditions~\cite{Spalding2019, Julia2022}. In practice, all of our VUMPS calculations
of the extended Hubbard model use a single-site update with a two-site unit cell, and we constrain our states to conserve $U(1)$ particle number and spin projection symmetry~\cite{Zauner-Stauber2018topological, itensor}.  We also constrain our states to be in the $S_z=0$ symmetry sector.  Results of convergence with bond dimension are shown in the Supplemental Materials~\cite{SM}.

Fig.~\ref{fig:fig1} (a) shows the energy $E$ on two bonds $B_{i} \; (i = 1, 2)$ that are associated with a two-site unit cell, computed using VUMPS.  It shows clear signals of both phase transitions.  For each fixed $U$, energies $E_{1}$ and $E_{2}$ inside one unit cell are exactly equal to each other in the SDW phase when $V$ is small.  As $V$ is gradually increased, $E_{1}$ and $E_{2}$ split, which reflects the broken translational symmetry of the BOW phase. The phase boundary between SDW and BOW can be determined quantitatively by setting a small threshold, e.g., where $|E_{1} - E_{2}| \sim 10^{-5}$.  As $V$ is further increased, the smooth or sudden changes in the energy per bond characterize the second-order $(U = 5)$ or first-order $(U = 6)$ phase transitions from BOW to CDW.

We can use the two bonds in a unit cell to partition the infinite system into two half-infinite subsystems and compute the von Neumann entanglement entropy $\mathcal{S}_{i}$.  As shown in Fig.~\ref{fig:fig1}(b), in the BOW phase, $\mathcal{S}_{i}$ has different values computed using different partitionings.  This corresponds to the spontaneously dimerized phase of the spin chain and the $\mathbb{Z}_{2}$ degeneracy of the two types of polarization~\cite{Sandvik2004, Ejima2007}.  In sharp contrast, the entanglement entropies computed in different ways of partitioning have exactly the same value in the SDW and CDW phases.  Therefore, the point where entanglement entropies deviate from one another can be used to locate the BOW phase boundaries, which gives results consistent with those obtained from the two energies.

% \tcb{The first referee would like our discussion to be less technical.  
% I have re-read our discussion, and I find it difficult to see what we could eliminate.  Basically,
% I just don't agree with the referee.  So I have instead pushed back in the response.  But in general
% it's better to bow to the wisdom of a referee if possible.  Can you folks see material we can move to the SM?
% If not, can you re-read the material which follows and see if you can make some small edits- remove anything
% un-necessary and replace technical language with more qualitative descriptive terms
% $\cdots$  That way we can say we did something at least.

%% I suggest we identify the
%% details which are less essential and move them to a new section of the SM.  At the same time, add a
%% few `big picture' comments which indicate to a PRL audience the general importance/applicability
%% of what we are doing, broader connections to other ML methods/physics, etc.  See draft referee response.
%% }

\paragraph{Sampling--}
We obtain our emulated experimental data by sampling finite subsystems of the translational invariant states found by VUMPS.  To obtain a sample, we repeat the tensors of the unit cell, and sample the resulting subsystem as one would sample a finite MPS~\cite{White2009, Stoudenmire2010}.  More specifically, we start by tracing our system down to a single site and sampling from the resulting density matrix, projecting onto the local state that was found and iterating the procedure over the finite subsystem.
%% Specifically, 
The unit cell is repeated sixteen times, providing samples that correspond to a chain of length $L = 32$ sites. For each value of $U$ and $V$, $N_{s} = 50000$ samples are collected. The sampled spin-resolved occupancy is stored in a feature array $\textbf{x}$ of length $2L$, where even and odd entries represent the spin-up and spin-down occupancy for each lattice site.  

%\tcb{[EK: This paragraph can move to SM if we were short in space.]}
\paragraph{Spontaneous symmetry breaking considerations--}
The states found by VUMPS spontaneously break the $\rm SU(2)$ symmetry of the model, and the spin direction of the state found by VUMPS will depend on details of the optimization, such as the initial state. Therefore, getting multiple samples from the same state obtained by VUMPS can be biased by the arbitrary spin direction of the state. To reduce this effect, we apply a random local $\rm SU(2)$ spin rotation uniformly to each site of the state before we obtain each sample.

While the continuous $\rm SU(2)$ symmetry is restored when producing the samples, the discrete $\mathcal{Z}_2$ symmetry present in BOW and CDW phases is broken in the VUMPS wave function [see Fig.~\ref{fig:fig1} (a) and \ref{fig:fig1}(b)]. Even if the experimental procedure does not break the $\mathcal{Z}_2$ symmetry in the CDW phase (the wave function is the homogeneous linear combination of both configurations), each individual sample will reflect which of the two ground states it comes from. There, the symmetry can be explicitly broken by post processing the samples, i.e., by translating by one site those that do not share the same pattern. Furthermore, the use of an even number of lattice sites, together with open boundary conditions (as commonly done in experiments) will break the $\mathcal{Z}_2$ symmetry in the BOW phase.  Open boundary conditions effectively provide a pinning field in the kinetic energy, forcing the strong bonds to be adjacent to the edges of the lattice~\cite{Julia2022} (see~\cite{SM} for a demonstration using exact diagonalization on chains of finite length). Therefore, the conclusions obtained with our emulated projective measurements are applicable to experimental data without loss of generality.

\begin{figure*}
    % \centering 
    \hspace{-.2in}
    \includegraphics[width = 1.\textwidth]{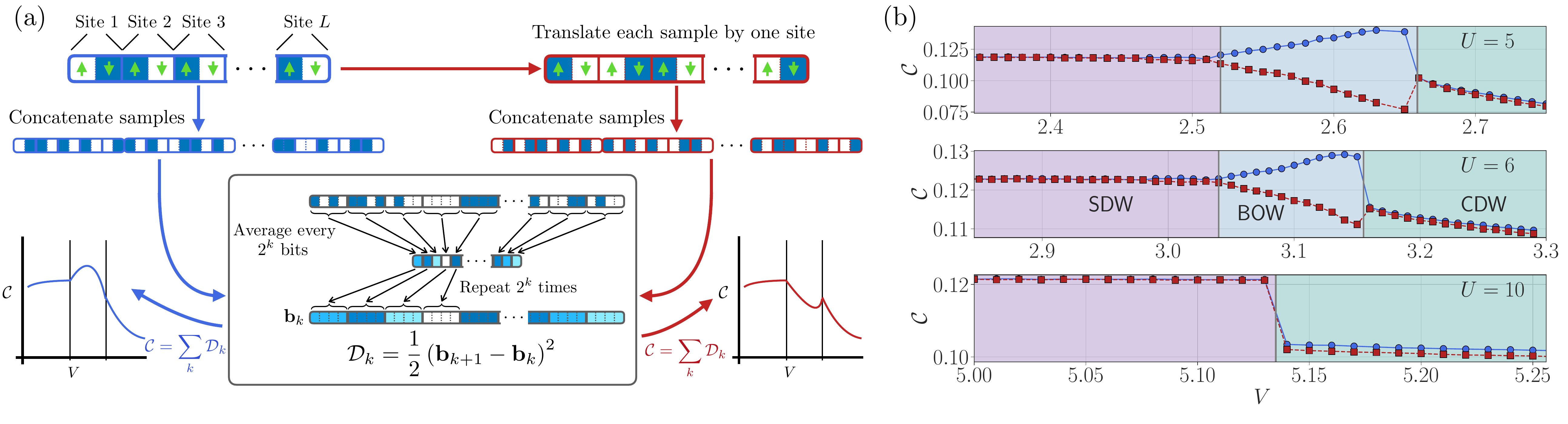}
    \caption{(a) Procedure for computing the structural complexity. $N_s$ samples are produced for a single ground state for a fixed value of $U$ and $V$. The samples are then: concatenated together, or translated by one lattice site and concatenated together. The multi-scale structural complexity is computed for both resulting bitstrings. One coarse-graining step is shown in the box, consisting in the computation of the average of adjacent groups of $2^k$ bits.  (b) Multi-scale structural complexity $\mathcal{C}$ as a function of nearest-neighbor interaction $V$ at fixed $U = 5$, $U = 6$, and $U = 10$.  The connected blue circles and red squares correspond to the upper and lower branches of the complexity measure in the BOW phase. The two branches are obtained by computing the structural complexity from the samples extracted directly from the emulated ground-state wave function and the samples obtained after translation by one lattice site as shown in panel (a).}
    \label{fig:fig3}
\end{figure*}

\paragraph{Principal component analysis.--}

Fixing the value of $U$, we run PCA on samples generated for different values of $V$ to explore fixed-$U$ cuts of the phase diagram. As a dimensional reduction method, PCA projects samples onto directions of largest variance in the data.
% \sout{computes the eigenvalues $\lambda_k$ and eigenvectors $\mathbf{w}^{(k)}$ of the covariance matrix of the samples. Then, the samples are projected onto the eigenvectors with the largest eigenvalues according to $y_k = \mathbf{x} \cdot \mathbf{w}^{(k)}$.}
It has been applied to detect phase transitions based on Monte Carlo samples for classical and quantum models~\cite{Wang2016-PCA-Ising, Wetzel2017, Hu2017, Khatami19}.

We find that the spread of the projected samples at fixed $V$ can be reduced, leading to a better resolution, if the input features $\mathbf{x}$ contain $N_{\textrm{c}}$ concatenated spin-resolved samples. 
Fig.~\ref{fig:fig2} (a) shows the relative weight of the eigenvalues of the covarience matrix of data, which represents the variance along the principal components, % \sout{$\lambda_k$} 
for samples generated at $U = 6$ and different values of $V$ for $N_\textrm{c} = 1$ and $N_\textrm{c} = 200$.  The increase in concatenated features reveals only one relevant principal component. The inset of Fig.~\ref{fig:fig2} (a) shows the average of the first principal component, 
% \sout{averaged over the indices of the concatenated samples $\big \langle \big \langle  \mathbf{w}^{(0)} \big \rangle\big \rangle$}
revealing its average action on the spin-resolved occupancy as the $\pi$ component of the Fourier transform of the total charge distribution. This quantity is the order parameter for the CDW phase.

 Fig.~\ref{fig:fig2}(b) shows the effect of the concatenation of the input features on their projection to the first two principal components. We observe that the first principal component resolves two clusters. The first one corresponding to samples in SDW and the BOW phases and the second one containing samples from the CDW phase. Fig.~\ref{fig:fig2} (c) shows, for a fixed value of $U$ and $V$, the average projection of the input features to the first and second principal components $(N_\textrm{c} = 200)$. As expected by the connection of the first principal component with the CDW order parameter, PCA can only resolve the BOW-CDW phase transition and its nature, % $1^\textrm{st}$ or $2^\textrm{nd}$ 
 first- or second-order (see~\cite{SM} for the PCA of the samples at different values of $U$). However, it shows no signal for the BKT-type transition between SDW and BOW phases.

\paragraph{Multi-scale structural complexity.--}

Recently, the multi-scale structural complexity measure~\cite{Bagrov20} has been used to obtain off-diagonal information about quantum states through projective measurements in a single basis~\cite{Sotnikov22}. As shown schematically in Fig.~\ref{fig:fig3} (a), the idea consists of concatenating all available samples for the same quantum state (creating a bitstring), performing several coarse-graining steps, and computing the dissimilarity $\mathcal{D}_k$ between consecutive coarse-graining steps $k$ and $k+1$ ~\cite{SM,Sotnikov22}. These dissimilarities are added except for the first step to obtain the so-called multi-scale structural complexity $\mathcal{C}$.

For each $(U, V)$ point, the $N_{s}$ spin-resolved samples are concatenated in two ways: (1) concatenation without shifting and (2) concatenation after translating all samples by one site (two bits with spin resolution) considering a periodic boundary for the bitstring.  These form two sets of bitstrings of length $2\cdot L\cdot N_{s}$ each.

As shown in Fig.~\ref{fig:fig3} (b), the multi-scale structural complexity captures
%% all 
both phase transitions.  The two sets of complexity analyses give essentially the same,
almost constant,
%% complexity
$\mathcal{C}$ (up to a constant shift) inside the SDW phase. 
%% which remains almost constant.  
As $V$ is increased, the transition to a BOW phase is clearly indicated by the splitting of the complexity measures into two branches. One branch 
%% of complexities 
increases as $V$ is increased while the other branch
%% of complexities 
decreases, corresponding to the two types of polarization of strong and weak kinetic energy bonds in the BOW phase.  The higher (lower) value of $\mathcal{C}$ is associated with a higher (lower) number of high kinetic energy bonds in the chain of length $L$.  As we keep increasing $V$, the two branches collapse into a single curve, indicating the transition to the CDW phase.  The absence of the BOW phase for $U=10$ is indicated by the lack of bifurcation of the complexity measure.

%As can be seen in Fig. \ref{fig:fig1} (b), this measure correlates very well with the entanglement entropy of the system, an observation first reported in Ref.~\cite{Sotnikov22} for Schr\"{o}dinger cat states. Like $\mathcal{S}$, the complexity tends to be smaller inside the CDW phase as compared to the SDW phase. 
It is worth noting that if we generate samples with equal probability from degenerate states in the BOW phase, the complexity does not bifurcate in the BOW phase like it does in Fig.~\ref{fig:fig3} (b). This is shown in the Supplemental Materials~\cite{SM}. It is then concluded that for the resolution of the BKT-type transition from the complexity analysis of single-basis projective measurements, we need samples that come from only one of the degenerate ground states. As discussed above, this can be achieved by imposing open boundary conditions~\cite{Julia2022, SM} or diagonal edge pinning fields~\cite{Assaad2013, Spalding2019, Xiao2023}.

\paragraph{Phase diagram--}
Fig.~\ref{fig:fig1}(c) compares the phase boundaries determined by the entanglement entropy $\mathcal{S}_{i}$ (red triangles) with those determined using the structural complexity $\mathcal{C}$, computed from samples directly (blue squares). The complexity analysis gives accurate results and quantitatively agrees with the off-diagonal observables computed from the wave function in the TDL within error bars.  Our results are also consistent with previous works~\cite{Sandvik2004, Ejima2007}.  Furthermore, obtaining the ground-state wave function in the TDL and combining local observables with machine learning approaches can shed light on the challenges of quantitatively locating the BKT-type transition~\cite{Sandvik2004, Zhang2004, Dalmonte2015, Spalding2019, Julia2022}.

\paragraph{Conclusion.--} 
In this work, we use the VUMPS algorithm to generate the 1D ground-state wave-function of the extended Hubbard model directly in the TDL, which allows us to determine phase boundaries with high precision without considering boundary effects and finite-size scaling.
% using local observables such as energy and von Neumann entanglement entropy without using multi-point functions and performing finite-size analysis.
% \tcg{[MF: This seems a bit strong to me, you can still use von Neumann entropy and local energy expectation values for finite systems, though it would have the disadvantage that you would have to do scaling analysis in both the system size and bond dimension. It seems like we should emphasize more that it allows us to ignore boundary effects and finite size scaling.]}
We sample real-space snapshots of finite length and use them along with
unsupervised learning methods to characterize the BKT-type phase transition between SDW and BOW phases as well as the first-order and second-order phase transition between BOW and CDW phases.  We find that off-diagonal long-range order cannot be detected by the PCA even after concatenation of samples.  However, using the structural complexity analysis, the off-diagonal long-range order can be detected using spin-resolved fermion density snapshots if these snapshots are generated from one of the degenerate ground states of the BOW phase. We argue that in optical lattice experiments, this can be achieved by imposing open boundary conditions. Our results indicate the potential of machine learning techniques in revealing microscopic mechanisms and hidden orders using projective measurements of corresponding thermal density matrix in quantum gas microscopes. 
While detection of phases with off-diagonal long-range order using diagonal descriptors has been demonstrated, 
further work is required to see if ML methods such as multiscale complexity can also
differentiate between BKT and second order transitions from which they emerge.
Likewise, it should be noted that while
the structural complexity effectively locates transitions, it
does not directly yield the physical nature of the order.

\paragraph{Acknowledgement}
We are indebted to Andrew Millis for carefully reading the manuscript.  We would like to thank Benedikt Kloss, Anders Sandvik, Miles Stoudenmire, Simon Trebst and Steven White for insightful discussions.  Flatiron Institute is a division of the Simons Foundation. 
The work of E.K.~and R.S.~was supported by the
grant DE-SC-0022311, funded by the U.S. Department of Energy, Office of Science. D.S. was supported by AFOSR: Grant FA9550-21-1-0236.

\newpage
\bibliography{EHMML}

\newpage

\onecolumngrid

\newpage

{\centering

{\normalsize \bf Supplemental Materials: Extracting Off-Diagonal Order from Diagonal Basis Measurements}
% {\Large \bf Supplemental Materials: Unsupervised detection of off-diagonal order from diagonal basis measurements}

\vspace{0.5cm}

Bo Xiao$^1$, Javier Robledo Moreno$^{1, 2}$, Matthew Fishman$^1$ Dries Sels$^{1, 2}$, Ehsan Khatami$^3$, and Richard Scalettar$^4$\\

\vspace{0.1in}

{\it
$^1$Center for Computational Quantum Physics, Flatiron Institute,\\
162 Fifth Avenue, New York, New York 10010 USA\\
$^2$Department of Physics, New York University, New York, New York 10003, USA\\
$^3$Department of Physics and Astronomy, San Jos\'{e} State University, San Jos\'{e}, CA 95192\\
$^4$Department of Physics\sout{} and Astronomy}, University of California, Davis, CA 95616 USA\\
}

\vspace{1cm}

\twocolumngrid

\renewcommand{\thesection}{S\arabic{section}}  
\renewcommand{\thefigure}{S\arabic{figure}}

\setcounter{equation}{0}
\setcounter{figure}{0}

\section{Principal Component Analysis}
PCA computes the eigenvalues $\lambda_k$ and eigenvectors $\mathbf{w}^{(k)}$ of the covariance matrix of the samples. 
Then, the samples are projected onto the eigenvectors with the largest eigenvalues according to $y_k = \mathbf{x} \cdot \mathbf{w}^{(k)}$. 
In Fig. 2 of the main text, we show the normalized eigenvalues, the eigenvector corresponding 
to the largest eigenvalue, and projections of data onto the first two principal components.

\section{Correlation length and von Neumann entanglement entropy}
Tensor network techniques allow us to efficiently approximate the state of systems composed of many degrees of freedom with a manageable number of relevant ones. For matrix product states (MPS), the connected correlation function asymptotically decays exponentially with distance~\cite{Orus2014, Rams2018} 
\begin{eqnarray}
    C(r) \sim e^{-r/\xi}.
\end{eqnarray}
The correlation length can be computed using the transfer matrix $\mathcal{T}$
\begin{eqnarray}
    \xi = -\frac{1}{\log|\lambda_{2} / \lambda_{1}|},
\end{eqnarray}
where $|\lambda_{1}| > |\lambda_{2}| \geq |\lambda_{3}| \geq ... \geq |\lambda_{D^{2}}|$ are the eigenvalues of $\mathcal{T}$ and $D$ is the bond dimension~\cite{Orus2014}.

\begin{figure}[t]
    \centering
    \includegraphics[width = .5\textwidth]{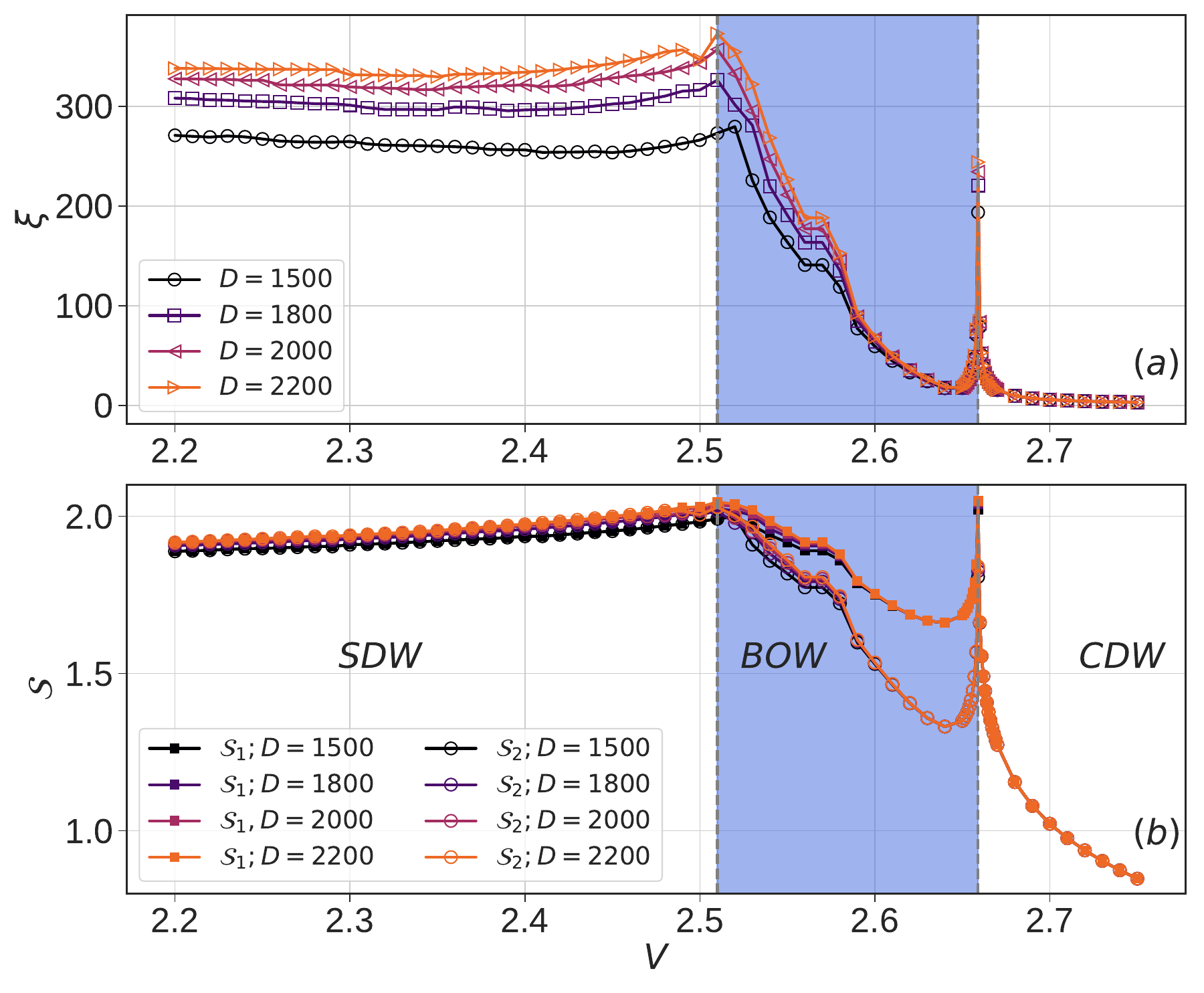}
    \caption{(a) Correlation length $\xi$ and (b) entanglement entropy $\mathcal{S}$ vs. nearest-neighbor interaction $V$ with fixed $U = 5$ using different bond dimensions.  The second-order phase transition from BOW to CDW is located at $V_{c2} = 2.659$, where sharp peaks in $\xi$ and $\mathcal{S}$ are observed.  The BKT-type transition from SDW to BOW is located $V_{c1} = 2.52$.}
    \label{fig:figS1}
\end{figure}

As shown in panel (a) of Fig.~\ref{fig:figS1}, the correlation length $\xi$ is large and grows with the bond dimension in the SDW phase (which is spin gapless) and diverges exactly at the continuous phase transition between the CDW and the BOW phase.  In sharp contrast, the correlation length $\xi$ has negligible dependence on the bond dimension and remains short distance in the CDW in which both the charge and spin excitation are gapped.  Inside the BOW, $\xi$ has clear dependence on the bond dimension near the BKT-type transition and gradually becomes bond dimension independent.

In addition to the correlation length $\xi$, the entanglement entropy is also a measure of correlations~\cite{Tagliacozzo2008, Pollmann2009, Pirvu2012}.  The von Neumann entropy of a pure state of a bipartite system AB is defined as,

\begin{eqnarray}
    \begin{aligned}
        \mathcal{S} &= -{\rm Tr} \rho_{A} \log \rho_{A} = -{\rm Tr} \rho_{B} \log \rho_{B}  \\
            % &= -\sum_{\alpha} \lambda_{\alpha}^{2} \log \lambda_{\alpha}^{2}.
    \end{aligned}
\end{eqnarray}
where $\rho_{A}(\rho_{B})$ is the reduced density matrix of subsystem A(B).  Inside a two-site unit cell, there are two bonds, along which we can divide an infinite chain into two half-infinite chains and then compute the entanglement entropy.  We denote the entanglement entropies computed by these two divisions $\mathcal{S}_{1}$ and $\mathcal{S}_{2}$ correspondingly, as shown in Fig.~\ref{fig:figS1}.

\begin{figure*}
    \hspace{-.2in}
    \centering
    \includegraphics[width = .9\textwidth]{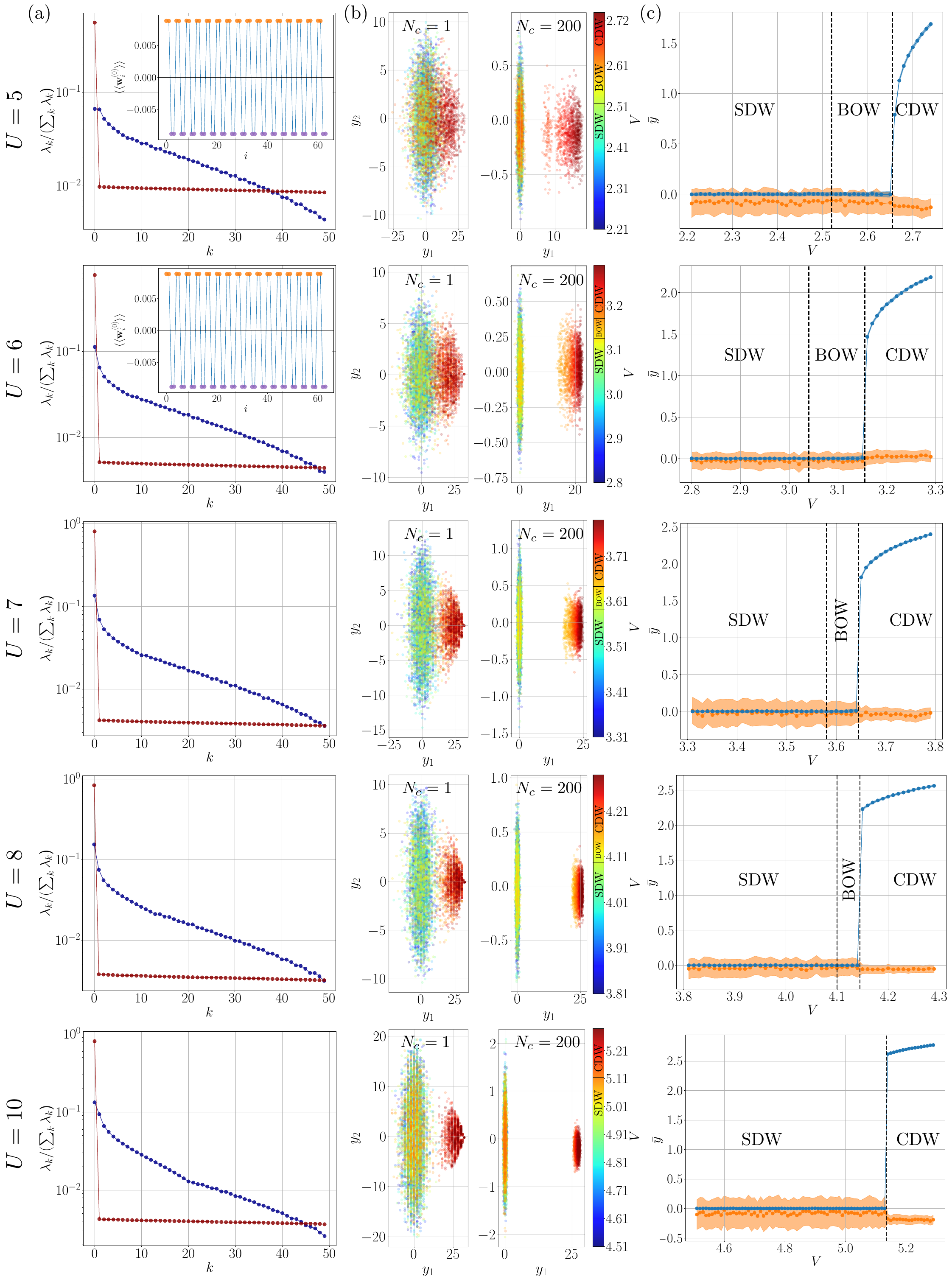}
    \caption{Principal component analysis of the samples generated at $U = 5, 6, 7, 8, 10$ as indicated on each row of figures and different values of $V$ on a uniform grid with separation $\Delta V = 0.1$. (a) Explained variance ratio measured by the relative weight of the eigenvalues of covariance matrix of data comparing two values of concatenated samples $N_\textrm{c}$. The inset shows the amplitude of the first principal component for $N_{\textrm{c} }= 200$, averaged over the concatenated samples. (b) Projection of the input features to the first two principal components both when $N_{\textrm{c} }= 1$  and $N_{\textrm{c} }= 200$. The color indicates the value of $V$ to which each projected feature belongs. (c) Average projection of the input features to the first (blue) and second (orange) principal components for $N_{\textrm{c} }= 200$. The shaded region shows the variance of the average value, measuring the spread of the projected input features at a fixed $V$. Vertical dashed lines indicate the phase transition points obtained form the entanglement entropy.}
    \label{fig:figS2}
\end{figure*}

In panel (b) of Fig.~\ref{fig:figS1}, we show how the von Neumann entanglement entropy $\mathcal{S}$ evolves along a vertical cut with $U = 5$ fixed on the phase diagram shown in Fig.~\ref{fig:fig1}(c).  The entanglement entropy peaks exactly at the BKT-type and continuous quantum phase transitions.  In the CDW phase, $\mathcal{S}$ decreases rapidly below $\mathcal{S} = 1$ as $V$ is increased, which indicates the system becoming more and more classical. Most importantly, given the broken translational invariance in the BOW phase, the two entanglement entropies deviate from each other, which reflects the $\mathcal{Z}_{2}$ degeneracy of two types of polarization.

\section{PCA for all $U$ values considered in this study.}
This section shows the results obtained from PCA at different values of $U$. Panel (a) of Fig.~\ref{fig:figS2} shows the variance ratio, defined as the relative weight of the eigenvalues of the covariance matrix, for $N_\textrm{c} = 1$ and $N_\textrm{c} = 200$. The effect of increasing the number of concatenated samples on the input features is to improve the resolution of the variance profile in the space defined by the samples. For $N_\textrm{c} = 200$, only one principal component is necessary to describe the variance properties of the data, for every value of $U$. As shown by the inset on panel (a) of Fig.~\ref{fig:figS2}, the first principal component is nearly identical for all values of $U$. As discussed in the main text, this particular profile for the first principal component computes the CDW order parameter for the samples. 

Fig.~\ref{fig:figS2} (b) shows the input features projected to the first two principal components for $N_\textrm{c} = 1$ and $N_\textrm{c} = 200$. In both cases, two clusters are resolved along the first principal component. The first cluster corresponds to samples that belong to SDW and BOW phases, while the second cluster contains mostly samples that belong to the CDW phase. The effect of increasing the number of concatenated samples of the input features is to provide a better distinction between the two clusters.

Panel (c) in Fig.~\ref{fig:figS2} shows, for five fixed values of $U$ and $V$, the average projection to the first and second principal components of the input features for $N_\textrm{c} = 200$.
%% , for different values of $U$ and $V$. 
For all values of $U$ the average projection as a function of $V$ remains featureless in the SDW and BOW phases, and increases rapidly in the CDW phase. It must be noted that this rapid increase is continuous at $U = 5$, while it is discontinuous at $U = 6, 7, 8, 10$, reflecting the second ($U = 5$) and first ($U = 6, 7, 8, 10$) order nature of the BOW-CDW transition. This observation comes as no surprise at the projection of the input features to the first principal component is analogous to the computation of the CDW order parameter.

\section{Structural complexity of samples with mixed polarizations}
\begin{figure}[h]
    \hspace{-.2in}
    \includegraphics[width = .5\textwidth]{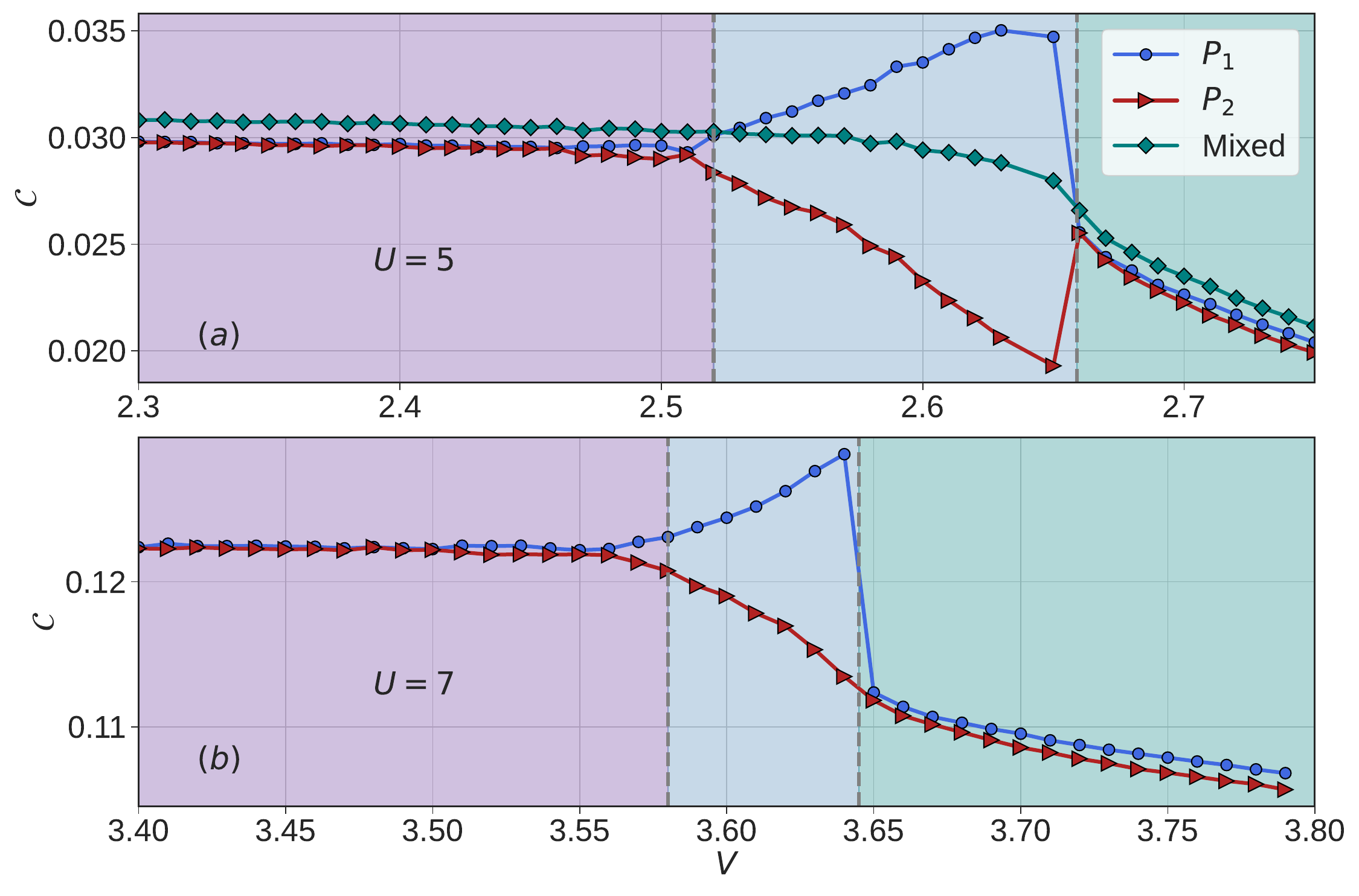}
    \caption{Multi-scale structural complexity $\mathcal{C}$ as a function of nearest-neighbor interaction $V$ at fixed (a) $U = 5$ and (b) $U=7$.  Blue circles and red triangles are results obtained by analyzing samples drawn from single ground state selected from the $\mathcal{Z}_{2}$ degenerate ground states.  Teal diamonds show results when samples are drawn from both of the degenerate states with equal probability.}
    \label{fig:figS3}
\end{figure}

In the main text, we show how to apply the structural complexity analysis to samples that are drawn from the same $\mathcal{Z}_2$ degenerate ground state.  When periodic boundary conditions (PBC) are used or the TDL is reached, sampling from one specific state among degenerate states is not possible.  In Fig.~\ref{fig:figS3}(a), we show results of the structural complexity when samples drawn from different degenerate ground states are mixed.  In the BOW and CDW states, we sample the two-fold degenerate ground-state wavefunctions simultaneously and select samples as the input data with equal probability.  As a result, the structural complexity is featurelesss near the quantum phase transition from SDW to BOW.  No kink or sudden change of slope is observed, compared to the case where samples with different polarizations are not mixed.  In contrast, the structural complexity is still able to capture the second-order transition from BOW phase to CDW phase, although the signal is weaker when samples are selected from both degenerate states.

As shown in Fig.~\ref{fig:figS3}(b), the details of how the lower branch of the structural complexity transits from the BOW phase to CDW phase as $V$ is increased depends on the nature of BOW to CDW transition.  For a second-order transition, e.g., at $U = 5$, the lower branch shows a dip before entering the CDW phase.  In contrast, if the transition is first-order, e.g., at $U = 7$, there is no dip appearing in the lower branch.

% \color{blue}
\section{Weak-coupling regime}
In the main text, our analysis centers on the intermediate- to strong-coupling regime, characterized by on-site interaction strengths of $U=5$, $U=6$, and $U=10$.  Drawing on the findings of previous studies \cite{Ejima2007, Spalding2019}, we posit that between $U=5$ and $U=6$, the system approaches a tri-critical point where the transition between BOW and CDW shifts from a second-order to a first-order transition.  At $U=10$, a direct first-order transition from SDW to CDW is anticipated, bypassing the BOW phase entirely.

\begin{figure}[h]
    \centering \hspace{-0.2in}
    \includegraphics[width=0.5\textwidth]{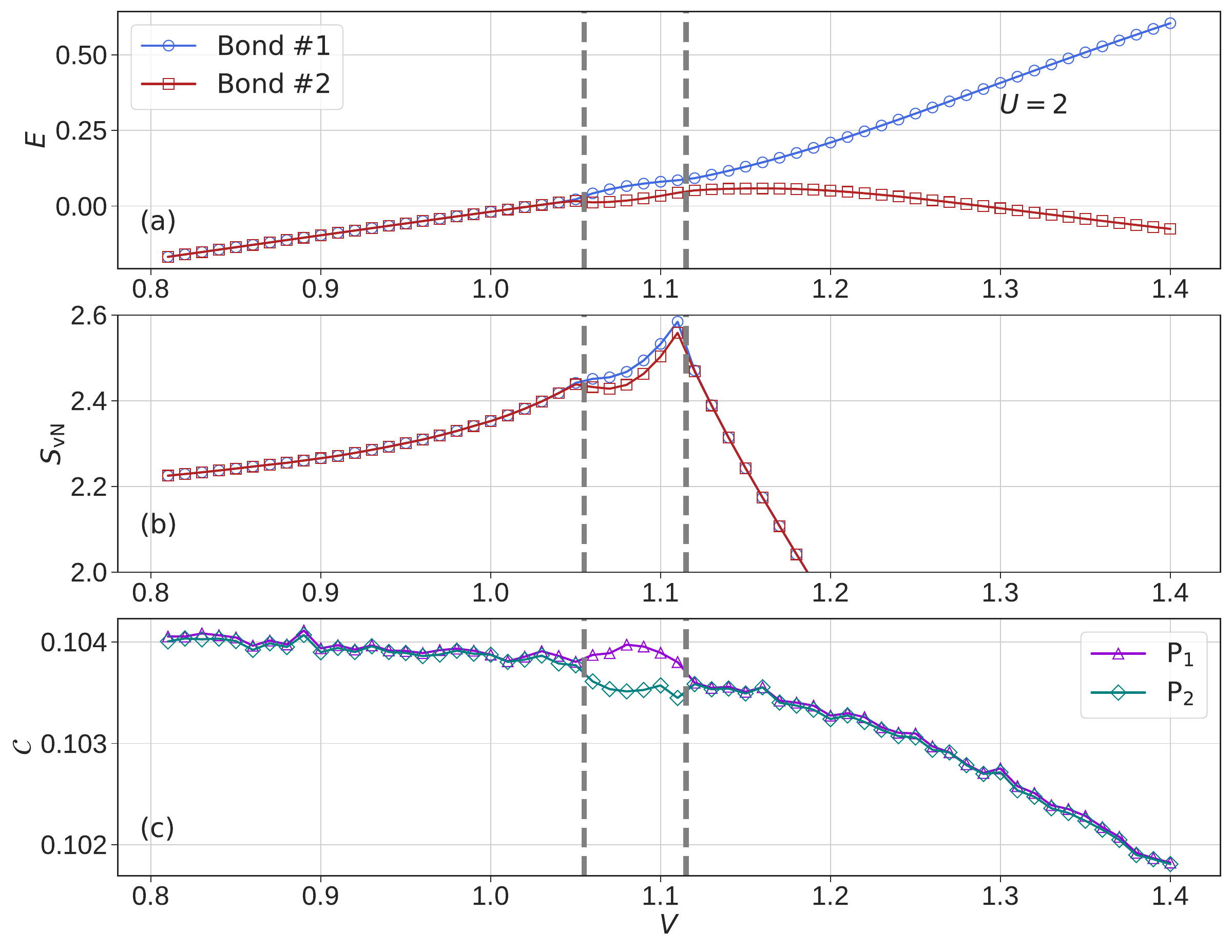}
    \caption{In the weak-coupling regime, we investigate: (a) the energy $E_{\ell}$, (b) the von Neumann entanglement entropy $S_{\rm vN}$ per bond for a single two-site unit cell at a constant on-site interaction strength of $U=2$, and (c) the complexity analysis of samples derived from sampling the ground-state wave function. 
    %The VUMPS algorithm exhibits outstanding performance in this regime, paralleling its efficacy in the strong-coupling regime, and provides precise insights into the two quantum phase transitions. Notably, the Bond-Order Wave (BOW) regime is considerably more constricted than in the cases with $U=5$ and $U=6$. Furthermore, the phase boundaries deduced from the structural complexity concur with those determined from VUMPS-derived metrics.
    }
    \label{fig:figS4}
\end{figure}

To validate the efficacy of our methodology, we have calculated the ground-state wave function at weak coupling, specifically for $U=2$.  As shown in Fig.~\ref{fig:figS4}(a)-(b),  $E_{B_{\ell}}$ and $S_{\rm vN}$ enable the identification of both BKT-type and second-order phase transitions with great precision. Fig.~\ref{fig:figS4}(c) shows that the structural complexity is again able to pinpoint both phase boundaries in this case. Hence, 
we find the VUMPS algorithm and the complexity analysis remain robustly effective even when, at $U=2$, the BOW phase is exceedingly narrow.
% \color{black}

%\textcolor{ROYALBLUE}{Utilizing the ground-state wave function, we can generate bitstring samples following the same procedure as described in the main text.  The spin-resolved samples are concatenated in two ways: concatenation without shifting and concatenation after translating all sites by one site, followed by applying the complexity analysis.  As shown in panel (c) in Fig.~\ref{fig:figS4}, the complexity $\mathcal{C}$ computed from these concatenated data sets coincides (up to a constant shift) in both the SDW and CDW phases.  Contrastingly, within the BOW phase, the complexity $\mathcal{C}$ exhibits distinct behaviors for the different polarizations, enabling the detection of the off-diagonal BKT-type transition with notable precision.  Moreover, it is remarkable how }

\section{Open boundary conditions and the BOW order}
Using exact diagonalization (ED) on finite systems we demonstrate that open boundary conditions (OBC) pin the bond order wave patterns in one dimensional chains as long as the number of sites of the chain $L$ is even.

\begin{figure}
    \centering
    \includegraphics[width = .5\textwidth]{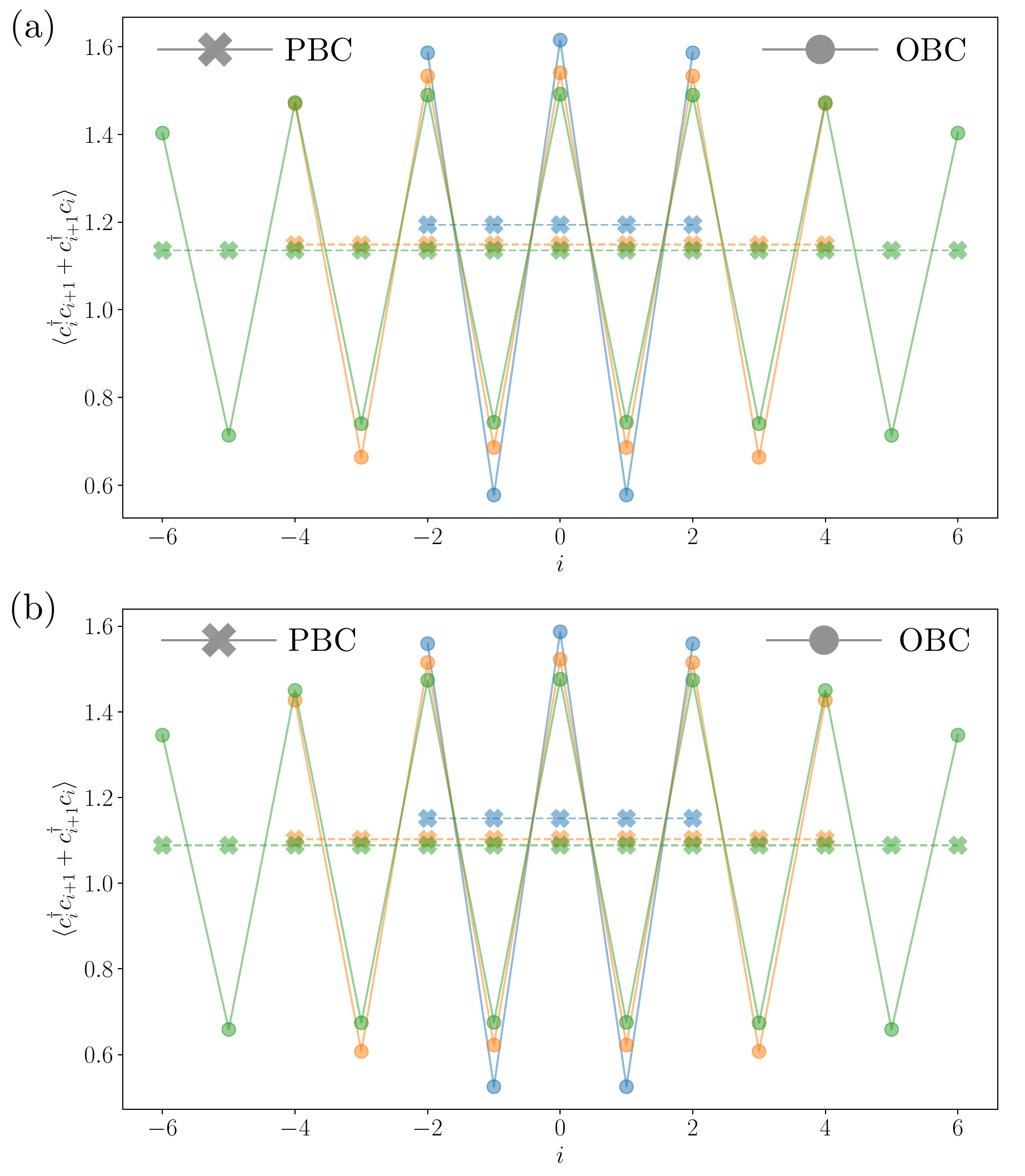}
    \caption{Kinetic energy
    between sites $i$ and $i+1$ 
    at each bond of the one-dimensional lattice, 
    both for open boundary conditions and periodic boundary conditions. Panels (a) and (b) correspond to $(U = 5; V = 2.58)$ and $(U = 6; V = 3.08)$ respectively. Both points are inside the BOW phase. Different chain lengths are shown: blue ($L = 6$), orange ($L = 10$) and green ($L = 14$).}
    \label{fig:figS6}
\end{figure}

Fig.~\ref{fig:figS6} shows the kinetic energy on each bond between chain sites $i$ and $i+1$, both for open and periodic boundary conditions, at two points in the phase diagram belonging to the BOW phase. When the boundary conditions are chosen to be periodic, the ground state found by ED is the uniform mixture between the two $\mathcal{Z}_2$ degenerate ground states of alternating high and low kinetic energy bonds. This translates into a flat profile of kinetic energy across the chain. For open boundary conditions, alternating bonds show a pattern of alternating high and low kinetic energies. The edges of the chain always have high kinetic energy (strong bond).

This can be explained by the following argument: OBC are the pinning field of infinite amplitude for BOW. OBC can be understood as forcing the kinetic energy at both bonds just outside of the chain to be zero. If the phase of the system is that of alternating high and low kinetic energies on neighboring bonds, it is easy to convince oneself that for even $L$, the kinetic energies at the end bonds of the chain have to be high. Furthermore,  at fixed particle number, the kinetic  energy is a semi-definite negative term in the Hamiltonian (it lowers the energy). For $L$ even, the number of strong kinetic energy terms (negative terms) is maximized by having the bonds on both edges to have high kinetic energy. Therefore, that configuration of kinetic energy bonds is the one that lowers the energy, and therefore the true ground state of the system.

\newpage
\end{document}